\documentclass[%
 aip,
 amsmath,amssymb,
preprint,
]{revtex4-1}

\usepackage{graphicx}
\usepackage{dcolumn}
\usepackage{bm}

\usepackage[utf8]{inputenc}
\usepackage[T1]{fontenc}
\usepackage{mathptmx}
\usepackage{etoolbox}
\usepackage{subcaption}
\usepackage{ar}
\usepackage{xcolor}

\makeatletter
\def\@email#1#2{%
 \endgroup
 \patchcmd{\titleblock@produce}
  {\frontmatter@RRAPformat}
  {\frontmatter@RRAPformat{\produce@RRAP{*#1\href{mailto:#2}{#2}}}\frontmatter@RRAPformat}
  {}{}
}%
\makeatother
\begin{document}

\preprint{AIP/123-QED}

\title[Flapping Dynamics of a Horizontal Flexible Sheet]{Self-Induced Flapping Dynamics of a Horizontal Flexible sheet in a Uniform Flow}
\author{Kalyani Panigrahi}
\affiliation{Department of Mechanical Engineering, BITS Pilani Hyderabad Campus, India}

\author{Pardha S Gurugubelli}
\email{pardhasg@hyderabad.bits-pilani.ac.in}
\affiliation{Department of Mechanical Engineering, BITS Pilani Hyderabad Campus, India}

\author{Sabareesh G R}%
\affiliation{Department of Mechanical Engineering, BITS Pilani Hyderabad Campus, India}

\date{\today}

\begin{abstract}
The case of a conventional flexible sheet in a uniform flow has been of interest in understanding the underlying physics of passive coupled dynamics between a flexible structure and a flow field. Gravity is known to influence the flapping instability and post-critical dynamics.
Interestingly, the flapping instability and dynamics of a thin flexible structure have been investigated either by neglecting the effects of gravity or by considering gravity along the length/span of the sheet. 
This study experimentally investigates the self-induced and sustained flapping dynamics of a thin flexible sheet positioned horizontally, with gravity acting along its bending direction. To explore the coupled interplay between gravitational, aerodynamic, and structural effects on the onset of instability and post-critical flapping dynamics, wind tunnel experiments are conducted across a range of physical parameters of the flexible sheet, such as its length ($L$), and aspect-ratio ($\AR$) for different wind speeds. To further understand the effects of gravity, the flapping behavior of a vertically mounted flexible sheet with gravity acting along its span has been investigated, and comparisons have been drawn with its horizontal counterpart. It has been observed that gravity along the bending does not influence the onset of flapping instability. The horizontally mounted flexible structures exhibit higher flapping amplitudes and frequencies when compared to their vertical counterparts. The observations in this study have direct relevance in the field of smart propulsion and energy harvesting devices.

\end{abstract}

\maketitle
\section{Introduction}



A thin, flexible sheet, when placed in an external flow field with its leading edge clamped, exhibits a flapping motion for flow speeds above a critical value. The critical flow velocity depends strongly on the competing stabilizing and destabilizing effects which depend on the sheet parameters such as length ($L$), thickness ($h$), aspect-ratio ($\AR$), bending stiffness ($K$), and fluid flow parameters such as free-stream velocity ($\overline{U}_0$) and viscosity ($\mu$). In addition to the fluid and structural parameters, another parameter that is known to impact the flapping dynamics is the direction of gravity ($g$) with respect to the orientation of the flexible structure. The coupled self-induced instability has been of interest for applications ranging from bioinspired propulsion to energy harvesting, biomedical engineering to aerospace. There have been multiple experimental, numerical, and analytical works that have investigated the effect of different fluid-structure parameters on the onset of the coupled instability as well as the post-critical flapping kinematics and dynamics.

Experimental works of \cite{taneda1968waving,datta1975instability,bejan1982meandering,zhang2000flexible,yamaguchi2000flutter,shelley2005heavy,lemaitre2005instability,watanabe2002experimental} investigated the effects of length ($L$) on the flapping behavior, wherein they have observed that their exists a critical length above which the structure exhibits flapping for a given velocity and the critical length value decreases with increasing velocity. Although Zhang et al.\cite{zhang2000flexible} did not observe any significant changes in frequency w.r.t. length, the works of\cite{taneda1968waving, yamaguchi2000flutter,bejan1982meandering,pang2010flutter} have shown that the flapping frequency decreases with the sheet's length.
Watanabe et al. \cite{watanabe2002experimental} reported that the critical velocity increases with the flexural rigidity of the sheet, and the work of Shelley et al. \cite{shelley2005heavy} demonstrated that a heavier sheet is more prone to flapping than their lighter counterpart. Eloy investigated the aeroelastic instability of a Mylar sheet in a wind tunnel \cite{eloy2008aeroelastic} to investigate the effect of $\AR$ on the critical velocity. The critical velocity has been observed to decrease with $AR$, thus concluding that slender bodies are more stable than their counterparts with higher $\AR$. He also observed that for small aspect-ratios ($\AR\le1$) the hysteresis in the onset of flapping instability is very small and, inversely, for the sheet's larger $\AR$, the hysteresis greatly increases.  Eloy's work further attributed the onset hysteresis phenomenon to the inherent planeity defects of the flexible structure \cite{eloy2012origin}. Hiroaki et al\cite{hiroaki2024flutter}. Kumar et al. \cite{kumar2021dynamics} experimentally investigated the fixed-point stable and post-critical flapping regimes using PIV and hot-wire measurements for a vertically supported thin flexible sheet attached at the $TE$ of a NACA0015 airsheet.
In addition to fixed-point stable and post-critical flapping states observed experimentally, the soap film experiments of Abderhmanne et al. \cite{abderrahmane2012nonlinear} reported a switching dynamical state wherein the structure continues to switch between a limit-cycle flapping state and a fixed-point stable state intermittently. Similar switching phenomena were also observed by Virot et al. \cite{virot2013fluttering} during the transition between two flapping modes.  

The flapping motion of a flexible structure develops a vortex street that is similar to von K$\acute{\mathrm{a}}$rm$\acute{\mathrm{a}}$n vortices, resulting in oscillatory drag and lift forces. The drag force, which is a combination of both pressure-induced and viscous skin friction, experienced by a flapping structure has been experimentally investigated in various works \cite{taneda1968waving,datta1975instability,durgesh2020experimental, virot2013fluttering, morris2009experiments}. These experimental works have shown that the drag coefficient exhibits a sharp increase at the critical velocity \cite{taneda1968waving, morris2009experiments, virot2013fluttering} when the flexible structure loses its stability. 
\cite{morris2009experiments}
 observed that a flexible sheets with larger aspect-ratios ($
\AR$) exhibit a greater drag coefficient compared to their smaller 
$\AR$ counterparts.

In addition to the experimental work, various numerical and theoretical approaches have been used to predict the impact of the fluid-structural parameters affecting the flapping instability. The numerical studies of \cite{zhu2002simulation, farnell2004numerical, sawada2007fluid} replicated the experimental work of \cite{zhang2000flexible}. 
The two-dimensional numerical works of \cite{connell2007flapping,gurugubelliJCP,gurugubelli2015self} have shown three distinct dynamical regimes, namely: (i) fixed point stable, (ii) LCO, and (iii) aperiodic flapping as a function of the mass ratio ($m^*$) and non-dimensional bending rigidity. Analytical works of \cite{watanabe2002theoretical,argentina2005fluid,eloy2007flutter,eloy2012origin,gurugubelliJAM} focused on understanding the effects of structure to fluid mass ratio ($m^*$) and non-dimensional bending rigidity on the flapping behavior explicitly taking into account the finite span of the plate, by using the slender body theory and Galerkin models. In these works, it has been reported that a flexible sheet exhibits a higher mode of flapping for low $m^*$. 
The analytical approaches from \cite{shayo1980stability,huang1995flutter,gibbs2012theory} have investigated the effect of the span of a flexible structure on the stability of large $\AR$ panels by using linearized plate and potential flow theories along with a classical Lagrangian one-dimensional beam structural model to predict the linear flutter boundary for finite-sized rectangular plates.


As discussed above, there is abundant literature on the influence of key parameters such as the geometric dimensions ($\AR$ and $L$), bending stiffness ($K$), free-stream velocity ($\overline{U}_0$) on the onset and post-onset flapping mechanism either by neglecting the effect of gravity or by considering the gravity along the span of the flexible structure. Although the works of Yamaguchi \cite{yamaguchi2000flutter} and Hiraoki \cite{hiroaki2024flutter} have investigated the dynamics of a flexible plate-like structure placed horizontally in a flow field, they have not investigated the impact of gravity on the flapping motion. Hence, the effect of gravity acting along the direction of bending remains poorly explored. 
Additionally, almost all the works that have investigated the forces acting on the flexible structure have employed numerical methods. The aerodynamic lift forces acting on a flexible structure undergoing a flapping motion have not been experimentally investigated.


The objective of this work is to investigate the impact of gravity acting along the direction of bending on the flapping kinematics and force dynamics for a horizontally mounted flexible structure. A detailed comparative analysis of the onset and post-onset regimes is performed for the horizontally and vertically mounted flexible structures, where the gravity will act along the span of the structure. Post-critical flapping kinematics involves the investigation of the transverse tip displacement, frequencies, and flapping modes of the flexible structure with the aid of a high-speed camera by varying its length ($L$) and aspect-ratio ($\AR$). In addition to the kinematics, this paper presents an overview of the aerodynamic lift forces acting on the horizontal flexible sheet. 

This paper is broadly divided into three sections, where the first section includes assessing the flapping kinematics of the horizontal flexible sheet in detail by varying the $\AR$ and $L$ of the sheet. The second section delves into the aerodynamic lift coefficients ($C_L$) measured during the flapping of the horizontal flexible sheet for various $AR$ and $L$. The third section presents a comparative analysis between the response dynamics of the horizontal sheet with its vertical counterpart in the onset and post-onset regimes.

\section{Experimental setup and methodology}
\begin{figure}
\centering
\begin{subfigure}[b]{0.49\textwidth}
\centering
\includegraphics[width=\textwidth,trim={10 0 13cm 2cm},clip]{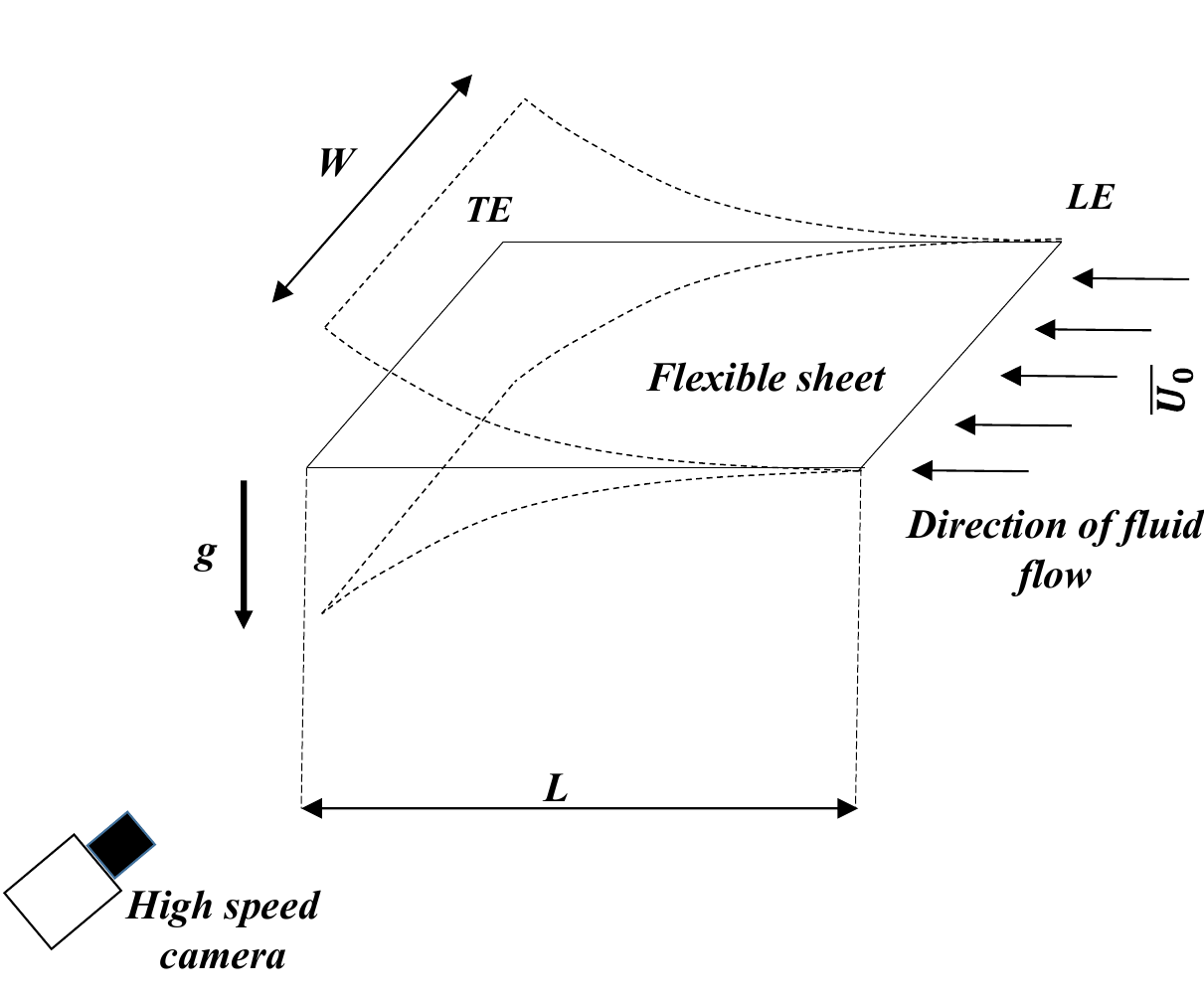}
\caption{ }
\label{fig:3d schematic-mod}
\end{subfigure}
\begin{subfigure}[b]{0.49\textwidth}
\centering
\includegraphics[width=\textwidth,trim={3.5cm 0 11cm 0},clip]{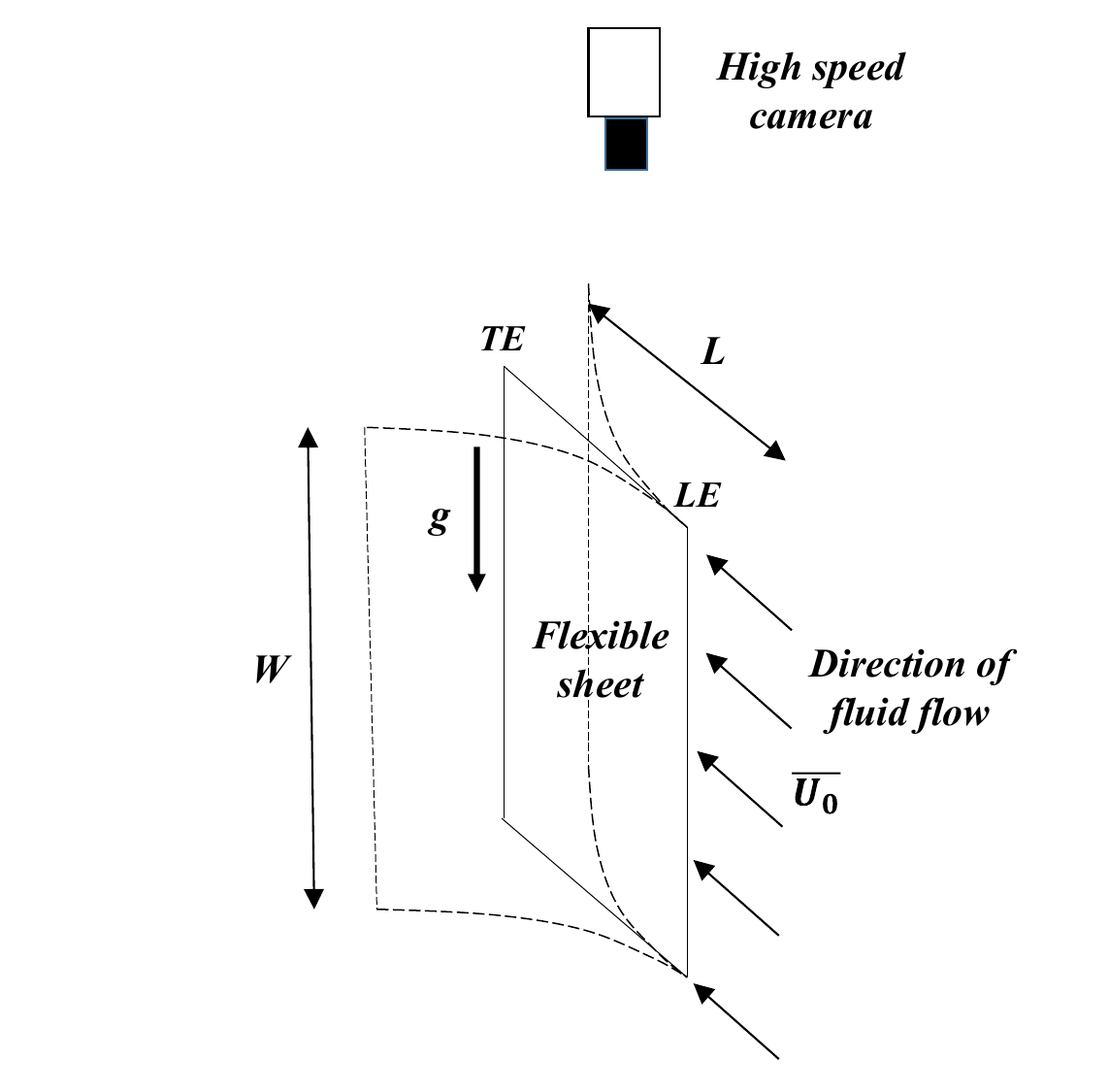}
\caption{ }
\label{fig:Schematics-vertical}
\end{subfigure}
\caption{ Comparing the three-dimensional representation of the flexible structure while it  flaps, (a) in a horizontal configuration 
 and (b) in a vertical configuration, (c) Top-view of the experimental setup inside the wind tunnel. }
\end{figure}
\begin{figure}
\centering
\begin{subfigure}[b]{0.85\textwidth}
\centering
\includegraphics[width=\textwidth,trim={1cm 2cm 0cm 2cm},clip]{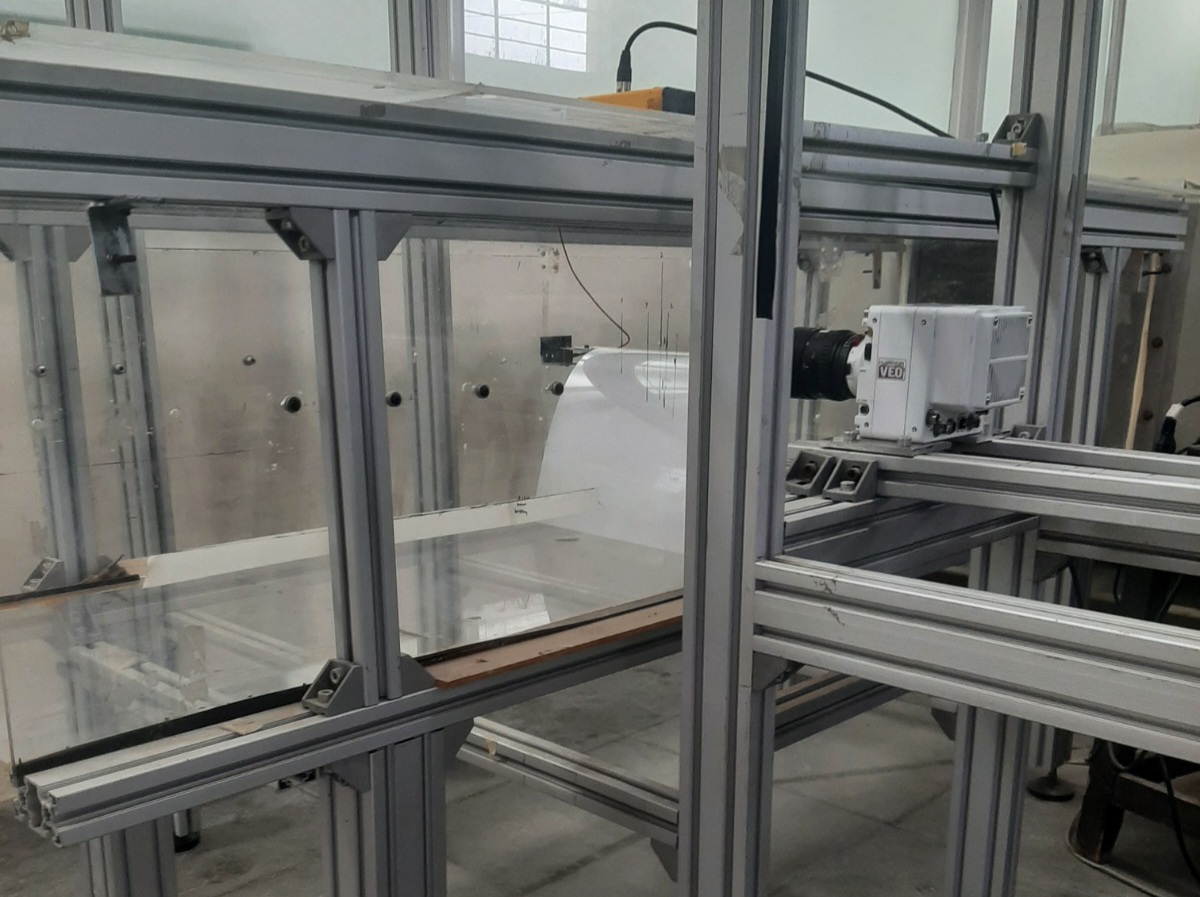}
\caption{ }
\label{windTunnel}
\end{subfigure}
\begin{subfigure}[b]{0.85\textwidth}
\centering
\includegraphics[width=\textwidth,trim={3cm 0cm 3cm 0},clip]{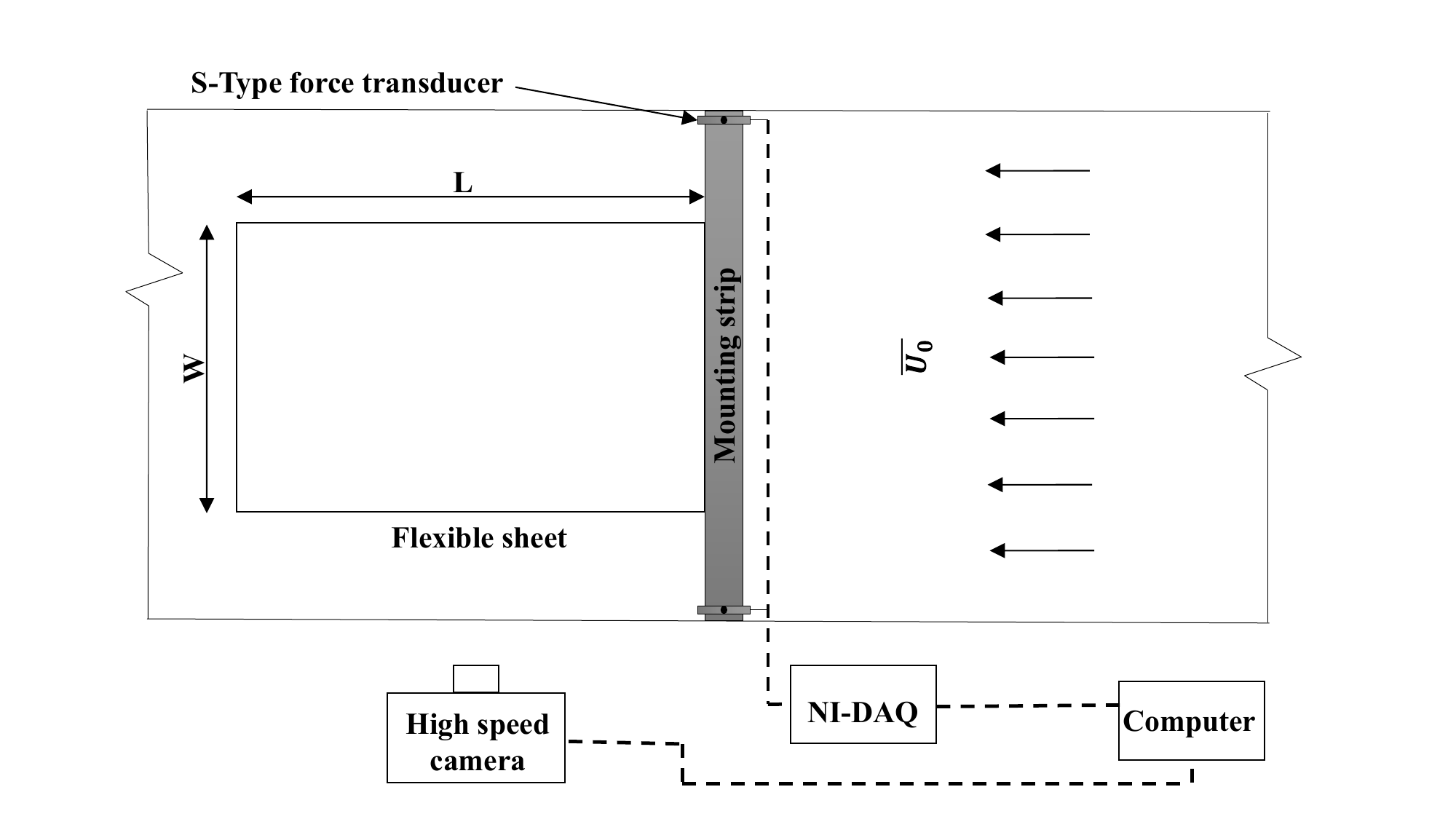}
\caption{ }
\label{fig:Schematics-1}
\end{subfigure}
\caption{The experimental setup inside the wind tunnel test section with the flexible structure mounted on the load cells.}
\end{figure}

To investigate the role of gravity on the flapping behavior of a thin flexible structure, a thin flexible Mylar sheet is placed horizontally in a uniform flow with gravity $g$ acting along the thickness/its direction of bending. The flexible sheet is clamped to its leading edge (LE) and the trailing edge (TE) is free to undergo a flapping motion as shown in Figure~\ref{fig:3d schematic-mod}. The thickness of the sheet considered is $h=0.178$ mm with a bending rigidity of $K =$$0.48~X~10^3$ Nm and density of $\rho\approx1400$ $\mathrm{kg/m^3}$. 
The experiments are performed in an open-ended low-speed wind-tunnel having a test section with a cross-sectional area of 47.5 x 47.5 sq cm and 2 m long as shown in figure~\ref{windTunnel}. The maximum airspeed that can be achieved inside the tunnel is about 12 m/s. The leading edge (LE) of the sheet is clamped onto a 4 mm thick aluminum strip with rounded corners positioned horizontally along the span of the test-section. The experiments have been carried out for free-stream velocities $(\overline{U}_0)$ $\le 6$ m/s so that the flapping kinematics remains largely two-dimensional. 

A Phantom high-speed camera (model: VE044L) is employed to capture the transverse tip displacement of the sheet ($\delta_y$) as shown in figure~\ref{fig:Schematics-1}. The high-speed imaging is captured at $200$ fps with a $2560~X~1600$ pixel resolution. Two light sources (Hiffin COB-120P with 3200-5200 $K$ color temperature range and GSVitech Multiled of 14000 Lumen intensity) are used for illuminating the entire edge of the flexible sheet. MATLAB-based edge detection and tracking techniques are applied to the high-speed images in processing the flapping amplitudes and frequencies. The strip carrying the flexible sheet is mounted on a pair of Futek miniature S-beam load cells, LSB205, of 3.92 N load capacity, for measuring the aerodynamic lift forces while the sheet flaps. NI-9237 bridge analog input module ($\pm$ 100 ppm accuracy, 24 bits ADC resolution, a maximum sampling rate of 50 kS/s for an input range of $\pm$ 25 mV/V) is utilized for recording voltage signals and measuring the aerodynamic lift ($C_L$ and $F_L$). Figure~\ref{fig:Schematics-1} shows a representative schematic of the experimental setup that includes sensors, DAQ, and a high-speed camera.

For a better understanding of the effect of gravity on the flapping kinematics between a sheet, a series of experiments are also performed by considering the gravity along the span, i.e., by placing a flexible structure vertically inside the wind-tunnel. 


\section{Results and Discussions}

Systematic parametric experiments have been conducted for a range of aspect-ratios ($\AR$) by varying the length ($L$) and width ($W$) for both the horizontal and vertical configurations, where $\AR=W/L$. For the case of horizontal configuration, the geometric parameters of the flexible sheet considered are $L \in \{22-28.5\}$ cm and $W \in [5-28.5\}$ cm. Similarly, for the case of vertical configuration, three different widths ($W$) are taken into consideration, $W~\in~\{28.5, 15, 5\}$ (in cm) for a constant length of $L =28.5$ cm with $\AR~\in~\{1, 0.52, 0.17\}$.
The response and force dynamics of the flexible sheet under the action of gravity have been compared between the horizontal and vertical configurations to investigate the influence of gravity on the flapping behavior.

\subsection{Horizontal configuration: Effect of gravity along the bending direction}\label{kinematics}
\subsubsection{Flapping kinematics}
\begin{figure}
\centering
\begin{subfigure}[b]{0.49\textwidth}
\centering
\includegraphics[width=\textwidth,trim={0.3cm 0 0.4cm 0},clip]{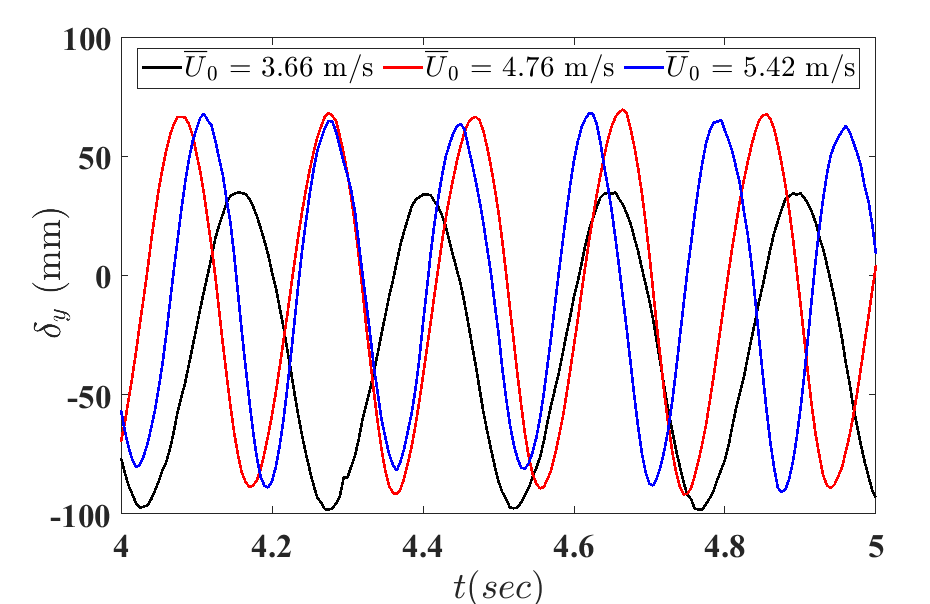}
\caption{ }
\label{fig:TipDisp_3diffvels}
\end{subfigure}
\hfill
\begin{subfigure}[b]{0.49 \textwidth}
\centering
\includegraphics[width=\textwidth,trim={0.3cm 0cm 0.4cm 0},clip]{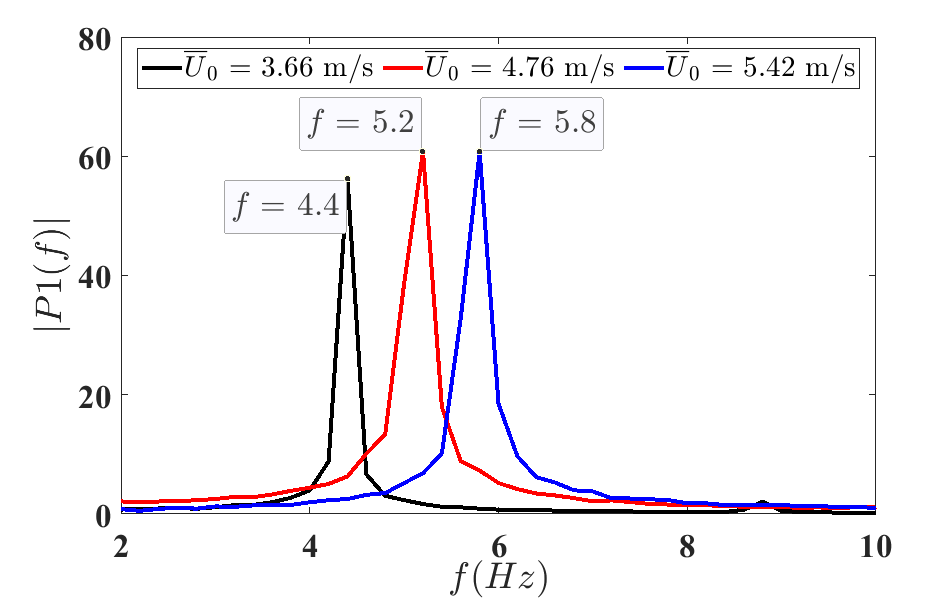}
\caption{ }
\label{fig:FFT_3diffvels}
\end{subfigure}
\caption{ (a) Time history of the $TE$ tip displacement ($\delta_y$) (b) Frequency power spectrum $|P1(f)|$ of the transverse tip displacement at $\overline{U}_0\in[3.66, 4.76, 5.42\}$ m/s for $L=28.5$ cm and $\AR=1$. } \label{effectU}
\end{figure}

Figure~\ref{effectU} shows the time history of TE's transverse displacements ($\delta_y$), about the fixed position of the LE, and the flapping frequency power spectrum for $\AR = 1$ and $L=28.5$ cm at three different wind speeds $\overline{U}_0 \in \{3.66, 4.76, 5.42\}~m/s$. It can be seen from the figure~\ref{fig:TipDisp_3diffvels} that the time series of the instantaneous sheet displacements clearly shows a sinusoidal, periodic, and uniform flapping motion with a clear dominant frequency depicting the limit cycle oscillation (LCO) regime. As the wind velocity is increased from $3.66$ m/s to $4.76$ m/s, both the flapping amplitude and frequency increase. Although there is an increase in the flapping frequency, no distinct change in the flapping amplitude is observed when the velocity is increased from $4.76$ m/s and $5.42$ m/s. The observed saturation of flapping amplitude following an initial rise with increasing velocity is consistent with the numerical findings reported in \cite{gurugubelli2015self}. This phenomenon can be attributed to the increase in the pressure drag experienced by the structure with increasing flapping amplitude. This saturation of the flapping amplitude marks the equilibrium between the pressure-induced flapping instability and stabilizing drag forces. 

\begin{figure}
\begin{subfigure}{0.7\textwidth}
\includegraphics[width=\columnwidth]{"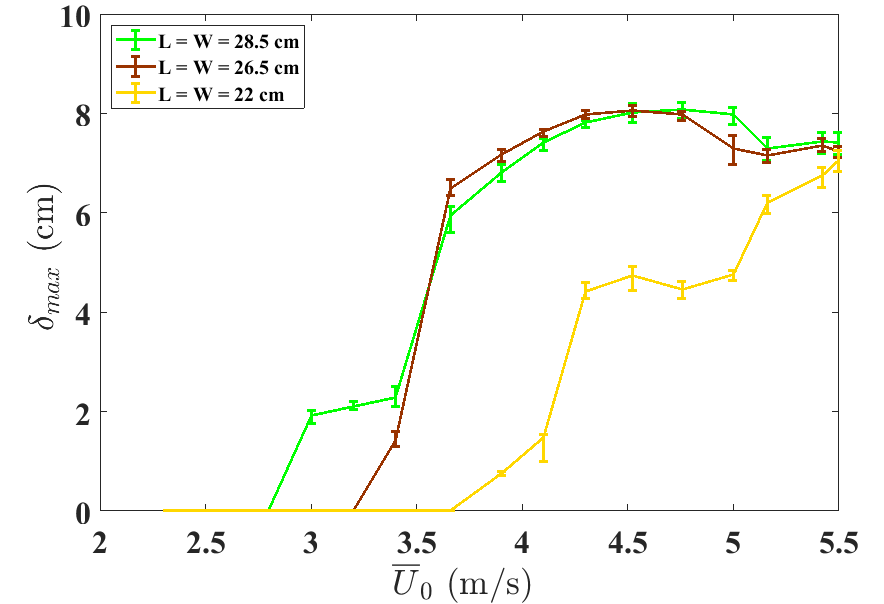"}
\caption{ }
\label{fig:maximum disp w.r.t length}
\end{subfigure}
\begin{subfigure}{0.7\textwidth}
\includegraphics[width=\columnwidth]{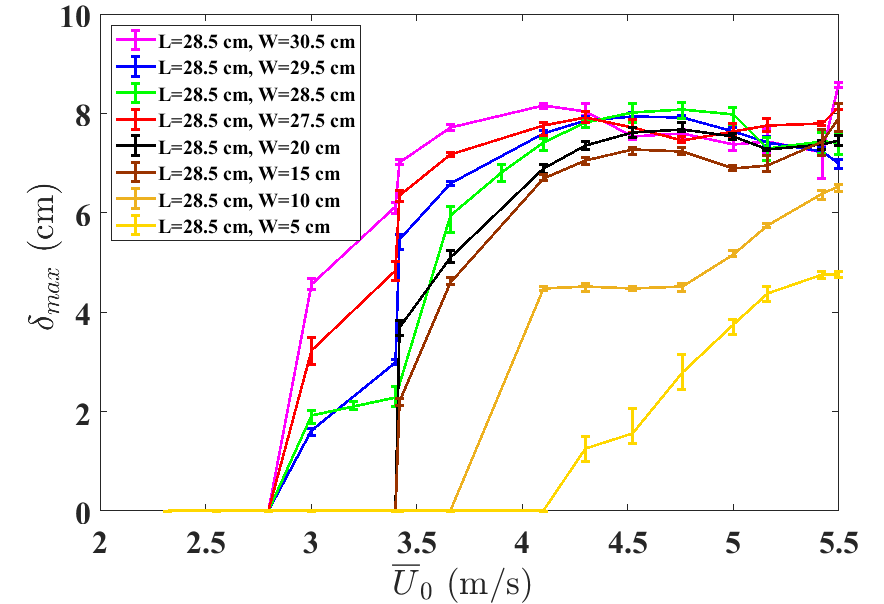}
\caption{ }
\label{fig:maximum disp w.r.t width}
\end{subfigure}
\caption{ Variation of maximum tip displacement ($\delta_{max}$) for (a) $L\in\{28.5,26.5,22\}$ cm for $\AR=1$ (b) $\AR\in[0.17-1.07]$ for $L=28.5$~cm. }
\end{figure}

To further understand the effect of velocity on the flapping phenomenon over a range of velocities and geometrical dimensions, figure~\ref{fig:maximum disp w.r.t length} depicts the variation of $\delta_{max}$ w.r.t. $\overline{U}_0$ for various lengths of the sheet $L \in \{28.5, 26.5, 22\}$ cm having a constant $\AR = 1$.
The shortest length considered here is 22 cm for $\AR=1$, as this is the critical length ($L_{cr}$) below which flapping stops.
There exists a critical velocity above which the flexible structure loses its stability and undergoes a flapping motion.
As the sheet loses its stability, a sharp jump in the flapping amplitude can be observed, followed by saturation of the flapping amplitudes. 
It can also be observed from figure \ref{fig:maximum disp w.r.t length} that as the length of the flexible sheet is reduced, the flapping onset is delayed. This observation can be attributed to the fact that as the flexible sheet becomes shorter, the bending stiffness of the sheet increases thereby making it more stable. Similar observations were also reported in the soap film experiments of \cite{zhang2000flexible} and in the numerical simulations of \cite{gurugubelliJAM}. 

Figure~\ref{fig:maximum disp w.r.t width} shows the variation of the $\delta_{max}$ as a function of widths ($W$), while maintaining a constant $L$. The $\AR$ for these sheets lies in the range of $\AR \in [0.17-1.07]$. A delay in the onset can be observed with decreasing $\AR$ even when the length is constant. 
A sudden jump in the transverse tip displacement ($\delta_{max}$) is also observed up to $\AR>=0.35$, but for $\AR<0.3$ the instability sets in more gradually. 
The $\AR$ effects are quite distinctively visible for $\AR<=0.5$. As can be seen from figure~\ref{fig:maximum disp w.r.t width}, wider flexible structures display larger flapping amplitudes compared to their lower $\AR$ counterparts, this observation can be attributed to the fact that as the $\AR$ reduces the side vortex impacts the pressure distribution on the top and bottom surfaces, thereby reducing the pressure difference between them and leading to lower flapping amplitudes. 
The onset of the instability becomes more gradual for lower $\AR$, because as the width ($W$) decreases, the mass per unit length and inertia of the structure reduce. From the literature \cite{connell2007flapping, gurugubelliJCP} it is well understood that when the inertia of the structure is low i.e. for lower mass-ratios the onset phenomena is more gradual when compared to their higher mass-ratio counterparts. 

\begin{figure}
\centering
\begin{subfigure}[b]{0.9\textwidth}
\centering
\includegraphics[width=0.70\textwidth,trim={1cm 5.5cm 13cm 6cm},clip]{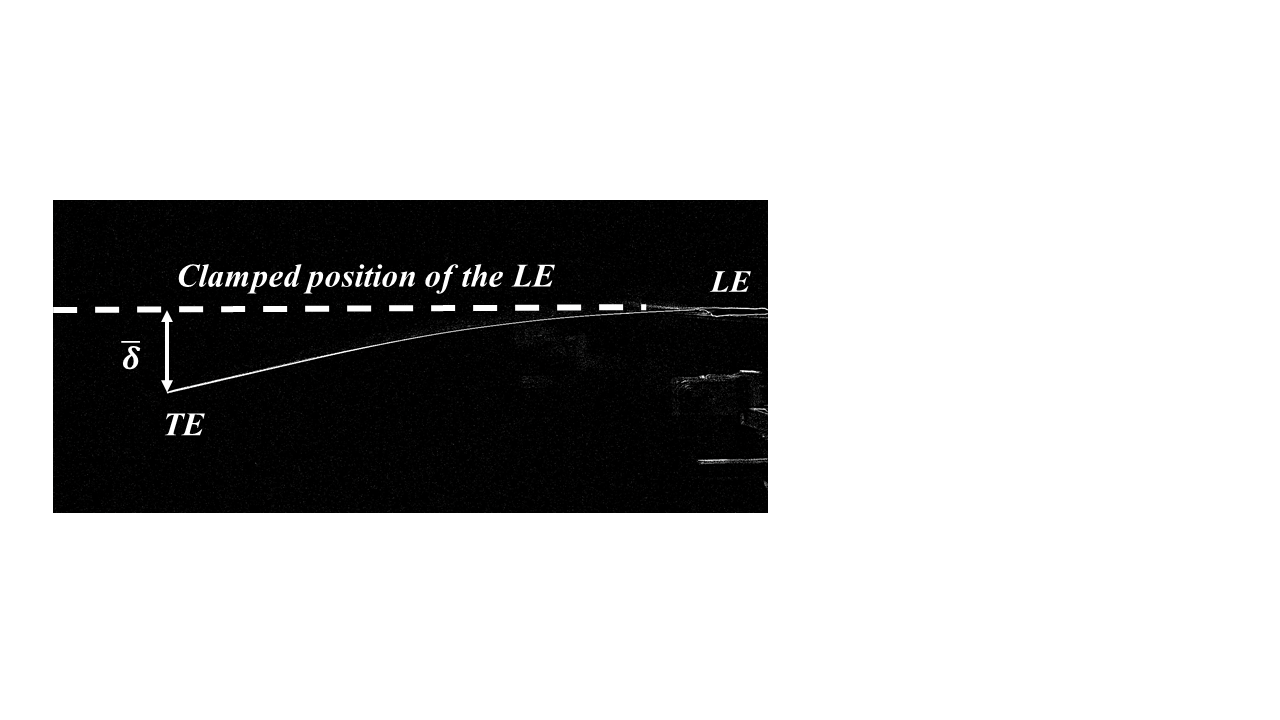}
\caption{ }
\label{schematicMeanDisp}
\end{subfigure}
\begin{subfigure}[b]{0.5\textwidth}
\centering
\includegraphics[width=\textwidth,trim={0.1cm 0cm 2cm 3},clip]{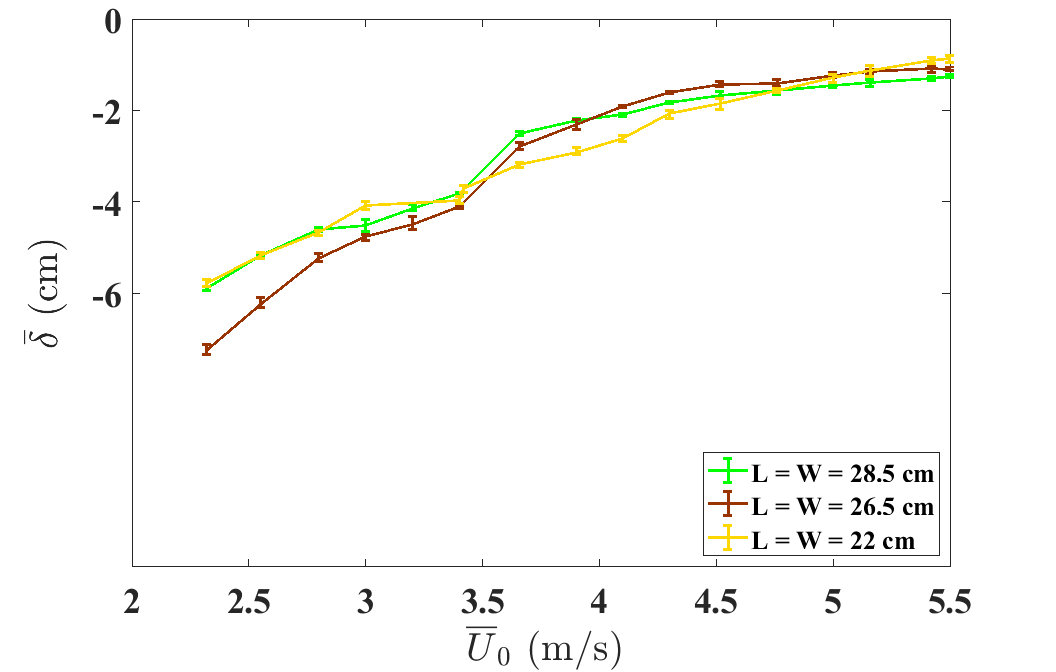}
\caption{ }
\label{fig:mean disp w.r.t length}
\end{subfigure}%
\hfill
\begin{subfigure}[b]{0.5\textwidth}
\centering
\includegraphics[width=\textwidth,trim={0.1cm 0cm 2cm 3},clip]{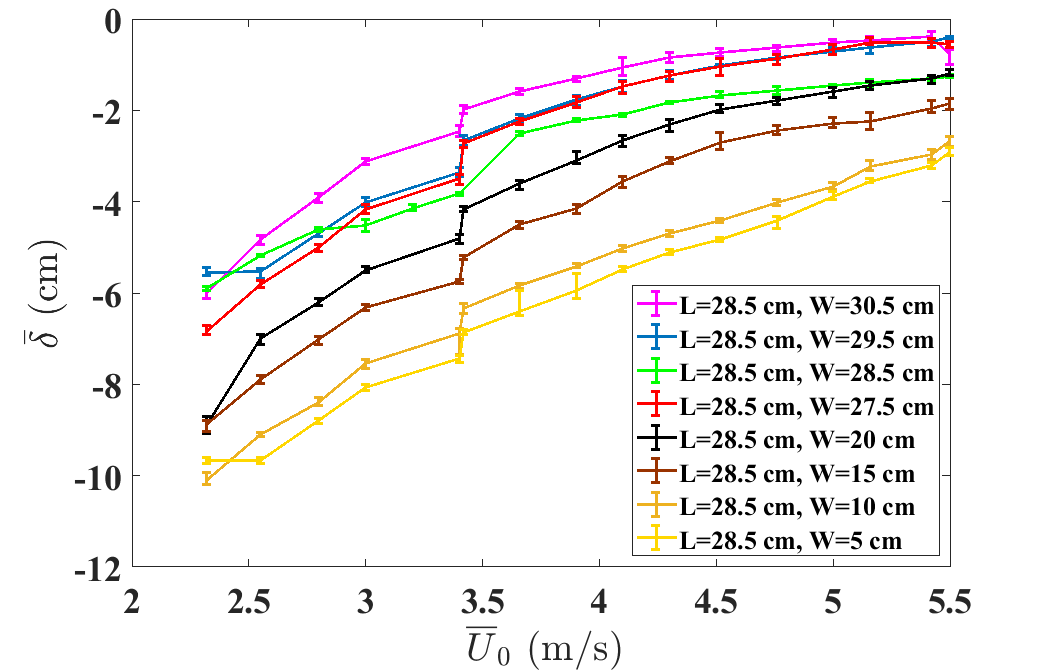}
\caption{ }
\label{fig:mean disp w.r.t width}
\end{subfigure}
\caption{ (a) High-speed camera snapshot of a flexible sheet $L = 22$ cm and $\AR = 1$ in a stationary state at $\overline{U}_0 = 3.66$ m/s showing the LE clamped position and the mean tip displacement $(\bar{\delta})$. Variation of $\bar{\delta}$ for (b) $L\in\{22-28.5\}$ cm and $\AR=1$ (c) $\AR\in\{0.17-1]$ and $L=28.5$~cm.}
\end{figure}

\begin{figure}
\centering
\begin{subfigure}[b]{0.49\textwidth}
\centering
\includegraphics[width=\textwidth,trim={0.75cm 0cm 1cm 0},clip]{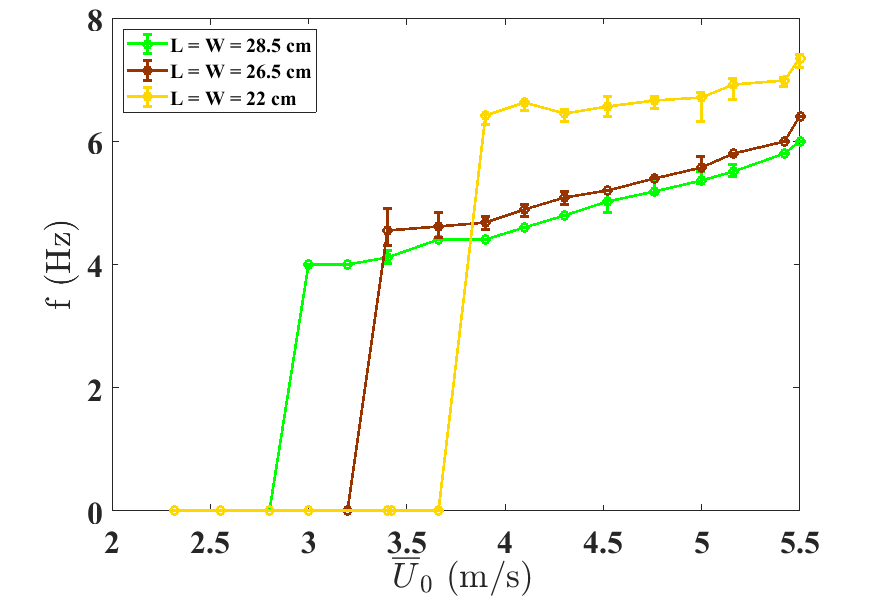}
\caption{ }
\label{fig:freq w.r.t length}
\end{subfigure}
\hfill
\begin{subfigure}[b]{0.49\textwidth}
\centering
\includegraphics[width=\textwidth,trim={0.75cm 0cm 1cm 0},clip]{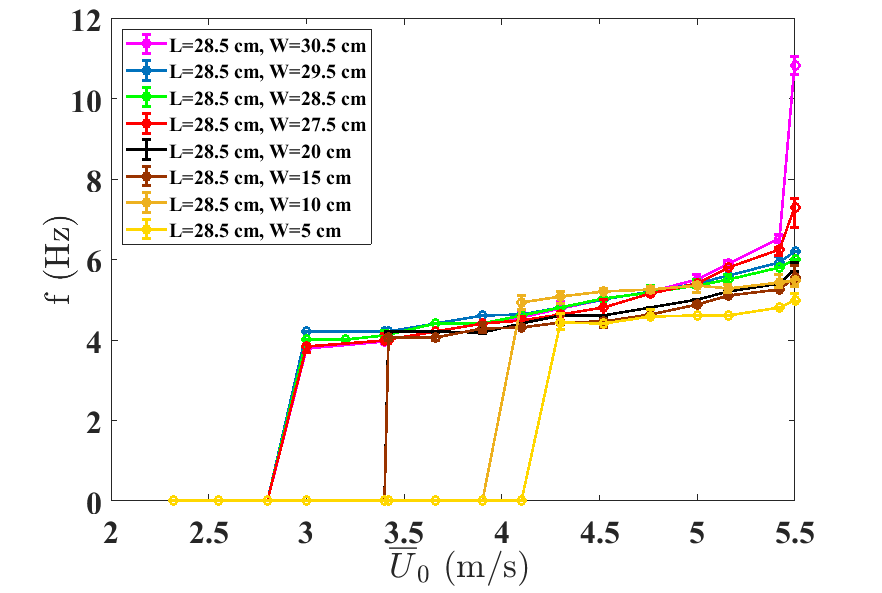}
\caption{ }
\label{fig:freq w.r.t width}
\end{subfigure}
\caption{ Variation of flapping frequency (f) for (a) $L\in\{22-28.5\}$ cm for $\AR=1$ (b) $\AR\in\{0.17-1]$ for $L=28.5$~cm. }
\end{figure}

Figures~\ref{fig:mean disp w.r.t length} and \ref{fig:mean disp w.r.t width} present the effect of $L$ and $\AR$ on the mean transverse tip displacement ($\bar{\delta}$) over a range of wind velocities, respectively. As the flexible sheet is placed in a flow field horizontally, the structure bends downward due to its weight. 
As the wind velocity is increased, the fluid dynamic drag acting on the structure lifts it up to an equilibrium position where there is a balance between the fluid dynamic forces and the weight of the structure. Figure~\ref{schematicMeanDisp} shows a high-speed camera snapshot corresponding to pre-onset stable regime and it can be observed that the flexible structure's TE lies below the clamped position of the LE. It has been observed that when the structure loses its stability for the post-critical regime, it now oscillates about the equilibrium position of the TE. With increasing velocity, the TE's equilibrium point would gradually move toward the LE's clamped position.    
From figure~\ref{fig:mean disp w.r.t width}, it can be seen that the sheets with lower $\AR$ bend downward more than their counterparts with higher $\AR$, this behavior can be attributed to the reduction in the pressure difference between the top and bottom surfaces of the structure
with reducing $\AR$ due to the side vortices \cite{gurugubelli2017thesis}. Furthermore, as $\AR$ decreases, the projected area in the flow direction reduces, leading to lower drag. Consequently, a flexible structure with a lower $\AR$ bends more in equilibrium than a wider one to compensate for the reduced projected area. 

Figures~\ref{fig:freq w.r.t length} and \ref{fig:freq w.r.t width} represent the flapping frequency (f) of the flexible sheet. From the figure~\ref{fig:freq w.r.t length}, it can be observed that the onset flapping frequency is significantly influenced by $L$, the flapping frequency increases with decreasing length. This behavior can be attributed to the increase in bending stiffness with a decrease in length. Additionally, from the figure one can also observe that the frequency gradually increases in the post-critical regime. From the figure~\ref{fig:freq w.r.t width}, it can be observed that the effect of $\AR$ on the post-critical flapping is not significant.

\begin{figure}
\centering
\begin{subfigure}[b]{0.325\textwidth}
\centering
\includegraphics[width=\textwidth,trim={0.2cm 0 1.5cm 0},clip]{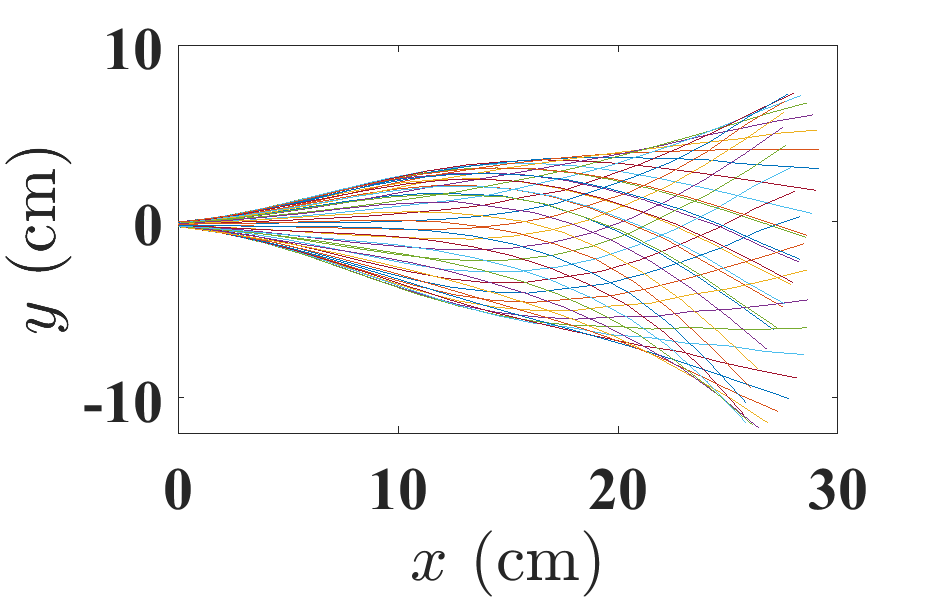}
\caption{ }
\label{fig:28.5by28.5V4.76_length}
\end{subfigure}
\hfill
\begin{subfigure}[b]{0.325\textwidth}
\centering
\includegraphics[width=\textwidth,trim={0.2cm 0cm 1.5cm 0},clip]{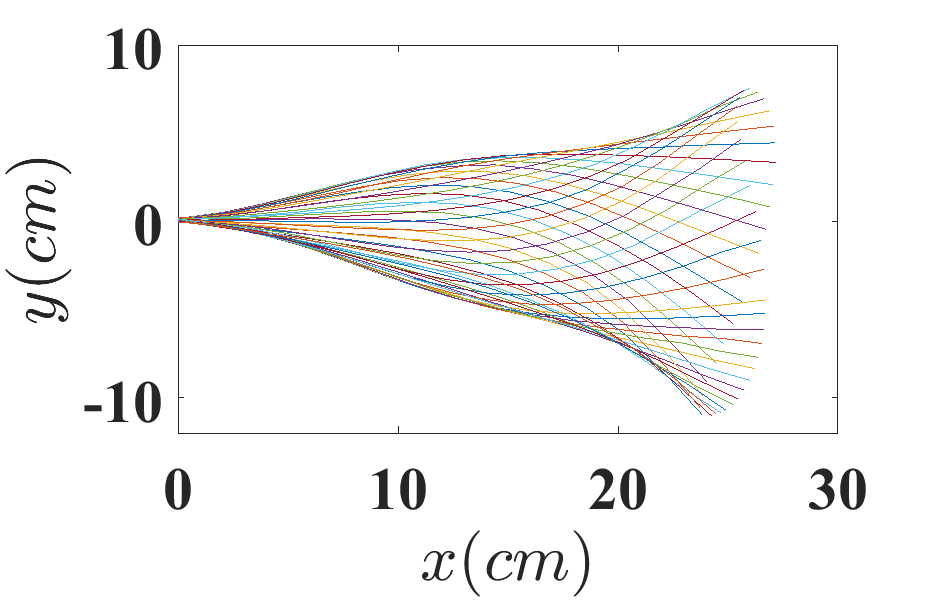}
\caption{ }
\label{fig:26.5by26.5V4.76}
\end{subfigure}
\hfill
\begin{subfigure}[b]{0.325\textwidth}
\centering
\includegraphics[width=\textwidth,trim={0.2cm 0cm 1.5cm 0},clip]{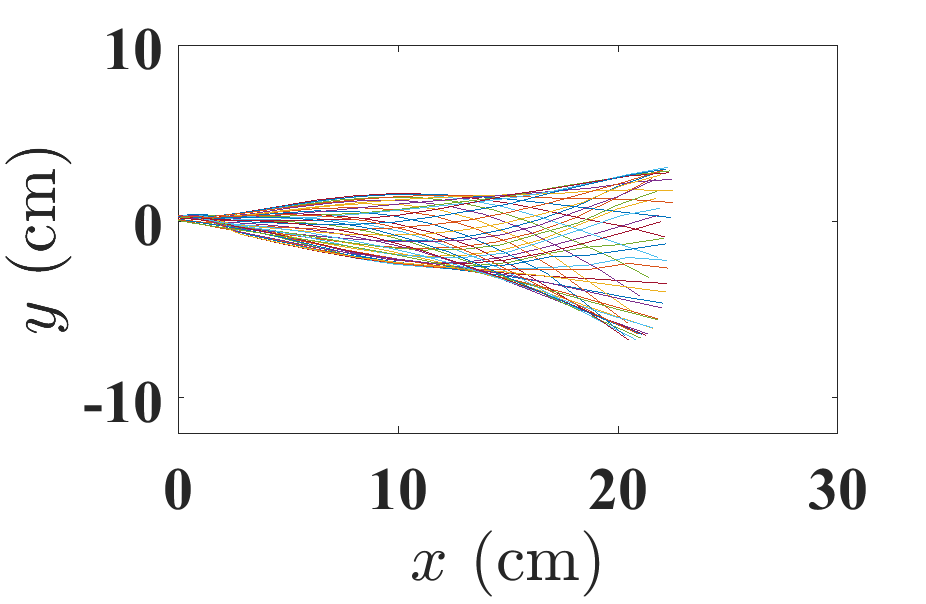}
\caption{ }
\label{fig:22by22V4.76}
\end{subfigure}
\caption{ Superimposed flapping modes of a flexible sheet over a flapping cycle with a constant $\AR=1$ for (a) L = 28.5, (b) L = 26.5, and (c) L = 22 at $\overline{U}_0$ = 4.76 m/s. }
\end{figure}

\begin{figure}
\centering
\begin{subfigure}[b]{0.325\textwidth}
\centering
\includegraphics[width=\textwidth,trim={0.2cm 0 1.5cm 0},clip]{Lengths_fb/28by28_4p76.png}
\caption{ }
\label{fig:28.5by28.5V4.76_width}
\end{subfigure}
\hfill
\begin{subfigure}[b]{0.325\textwidth}
\centering
\includegraphics[width=\textwidth,trim={0.2cm 0cm 1.5cm 0},clip]{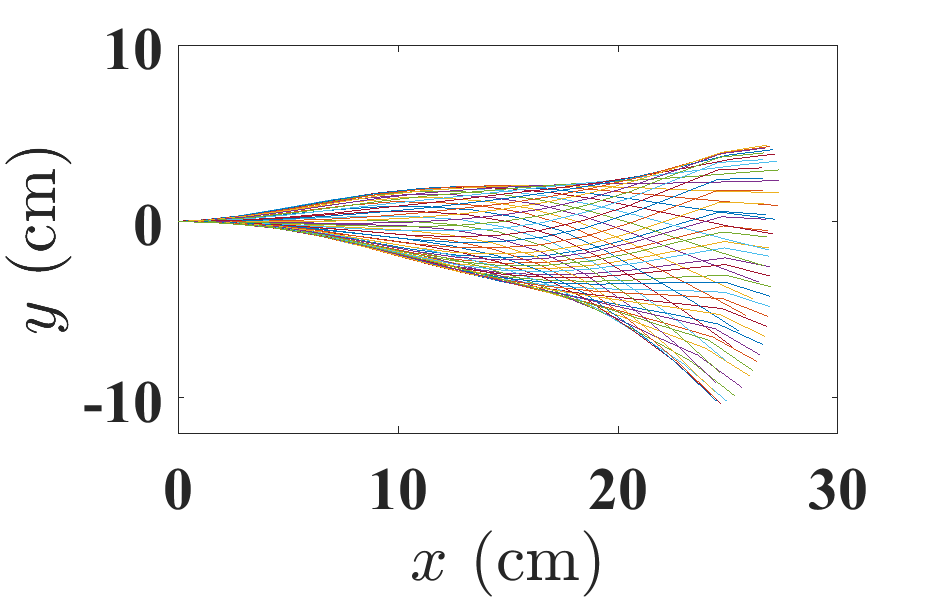}
\caption{ }
\label{fig:28.5by15V4.76}
\end{subfigure}
\hfill
\begin{subfigure}[b]{0.325\textwidth}
\centering
\includegraphics[width=\textwidth,trim={0.2cm 0cm 1.5cm 0},clip]{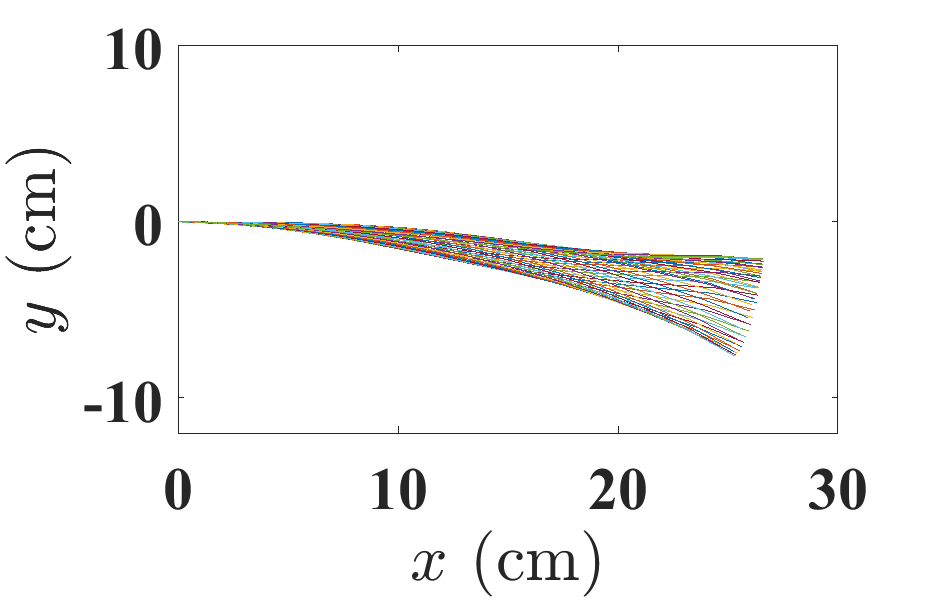}
\caption{ }
\label{fig:28.5by5V4.76}
\end{subfigure}
\caption{ Superimposed flapping modes of a flexible sheet over a flapping cycle with a constant $L=28.5$ cm for (a) $\AR$ = 1, (b) $\AR$ = 0.53, (c) $\AR$ = 0.17 at $\overline{U}_0$ = 4.76 m/s. }
\end{figure}

Figures~\ref{fig:28.5by28.5V4.76_length} to \ref{fig:22by22V4.76} present the mode shapes by superimposing the high-speed camera snapshots of the flexible sheet's edge over a flapping cycle for three different $L \in \{28.5, 26.5, 22\}$ cm, respectively, having $\AR$ = 1 at  $\overline{U}_0 = 4.76~m/s$ to understand the influence of $L$ on the post-critical flapping. It can be seen from the figure~\ref{fig:28.5by28.5V4.76_length} no significant change in the flapping is observed and only the flapping amplitude decreases with length. Additionally, the necking phenomenon shifts towards the LE with decreasing $L$. The flexible structure exhibits deformed flapping motion about a curved shape due to the bending of the structure under its own weight.
Figures~\ref{fig:28.5by28.5V4.76_width} to \ref{fig:28.5by5V4.76} present the effect of $\AR$ on the mode shapes for $L=28.5$ cm and W $\in \{28.5, 15, 5 \}$ cm at $\overline{U}_0 = 4.76~m/s$ for one oscillation cycle. From the figures, it can be observed for the lower $\AR$ the structure undergoes flapping motion fully below the LE clamped position.

\subsubsection{Flapping force dynamics}

\begin{figure}
\begin{subfigure}[b]{0.49\textwidth}
\centering
\includegraphics[width=\textwidth,trim={0.02cm 0cm 1.7cm 0.5cm},clip]{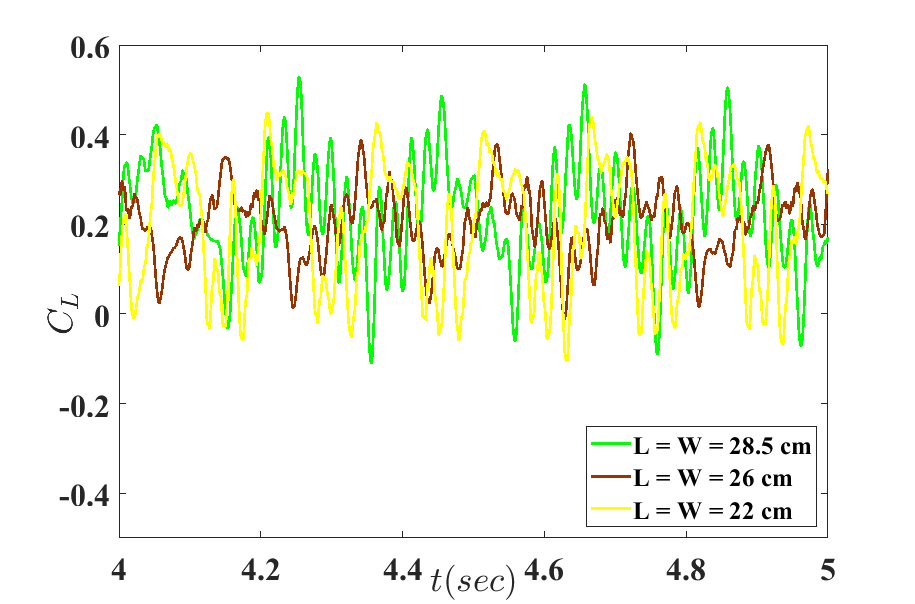}
\caption{ }
\label{fig:CL time history for all lengths}
\end{subfigure}
\begin{subfigure}[b]{0.49\textwidth}
\centering
\includegraphics[width=\textwidth,trim={0.2cm 0cm 1cm 0},clip]{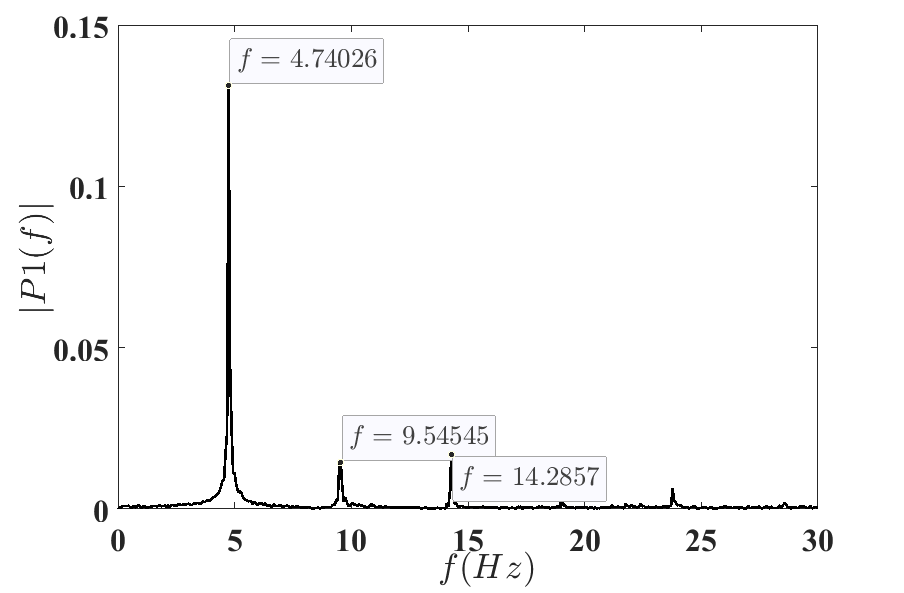}
\caption{ }
\label{fig:forcefft28by28}
\end{subfigure}
\begin{subfigure}[b]{0.49\textwidth}
\centering
\includegraphics[width=\textwidth,trim={0.5cm 0cm 1cm 0},clip]{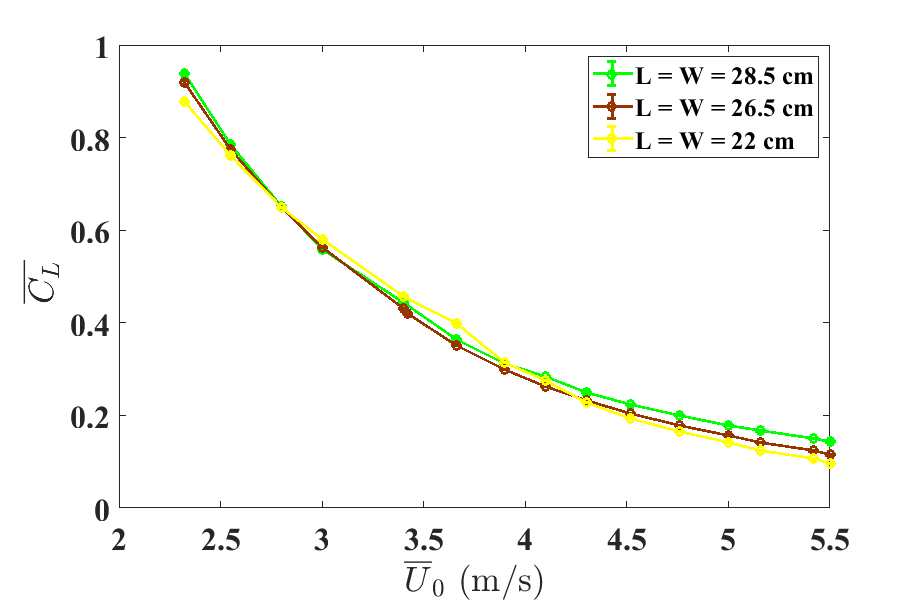}
\caption{ }
\label{fig: mean Lift coeff_summary for all lengths}
\end{subfigure}
\begin{subfigure}[b]{0.49\textwidth}
\centering
\includegraphics[width=\textwidth,trim={0.3cm 0cm 1cm 0},clip]{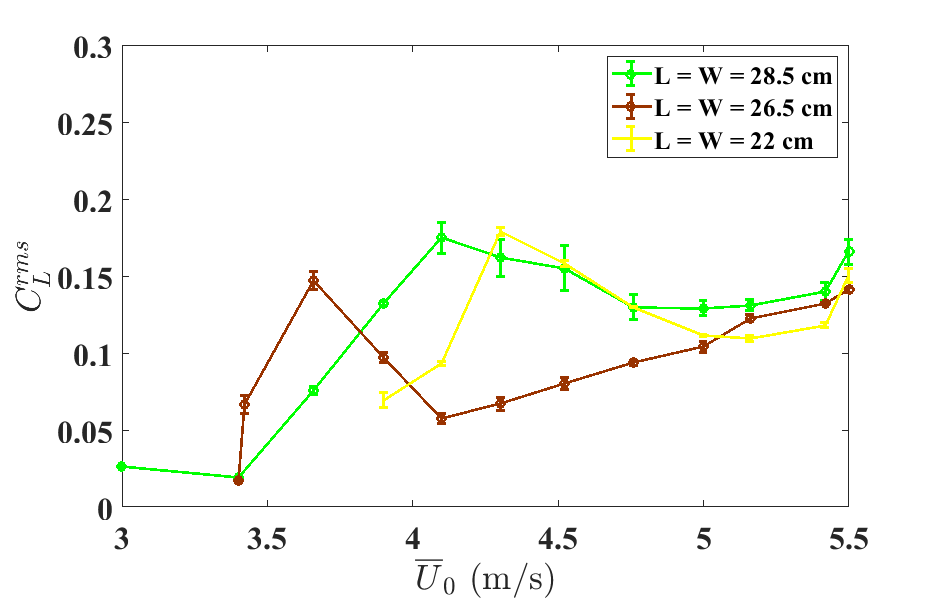}
\caption{ }
\label{fig:rms Lift coeff_summary for all lengths}
\end{subfigure}
\caption{ (a) Time history of $C_L$ for three different flexible structure's length $L=\{28.5, 26.5, 22\}$cm for $\AR=1$ and $\overline{U}_0 = 4.52$ m/s. (b) Power spectral distribution of the $C_L$ for $\overline{U}_0=4.52$, $L=28.5$ cm and $\AR=1$. Variation of: (c) mean $C_L$ ($\overline{C}_L$) and (d) Root mean square of $C_L$ (${C_L^{rms}}$) about the mean with $\overline{U_0}$ for $L \in \{28.5, 26.5, 22\}$ cm and $\AR=1$. } \label{forceConstLength}
\end{figure}
\begin{figure}
\centering
\begin{subfigure}[b]{0.49\textwidth}
\centering
\includegraphics[width=\textwidth,trim={0.02cm 0cm 1.8cm 0.5cm},clip]{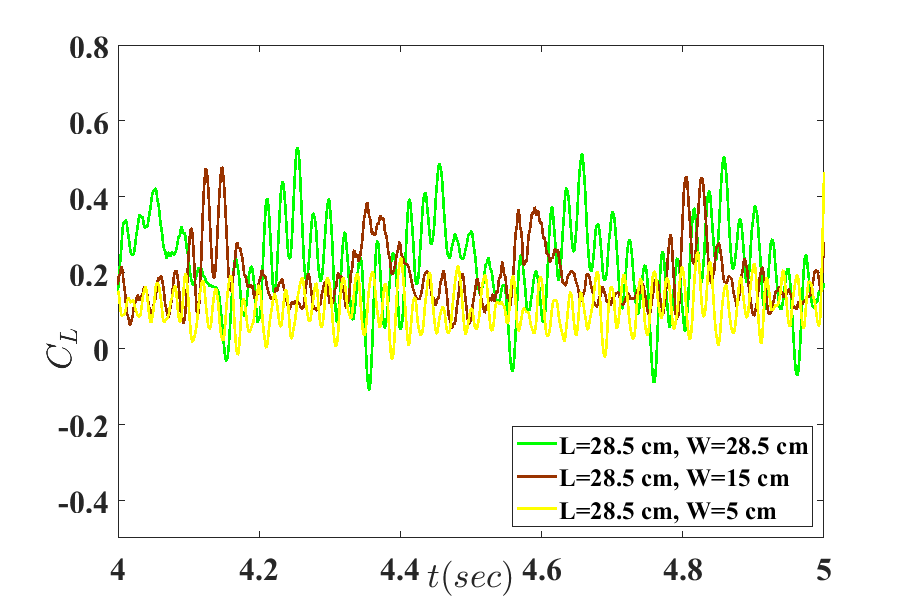}
\caption{ }
\label{fig:CL for all widths}
\end{subfigure}
\hfill
\begin{subfigure}[b]{0.49\textwidth}
\centering
\includegraphics[width=\textwidth,trim={0.5cm 0cm 1.5cm 0},clip]{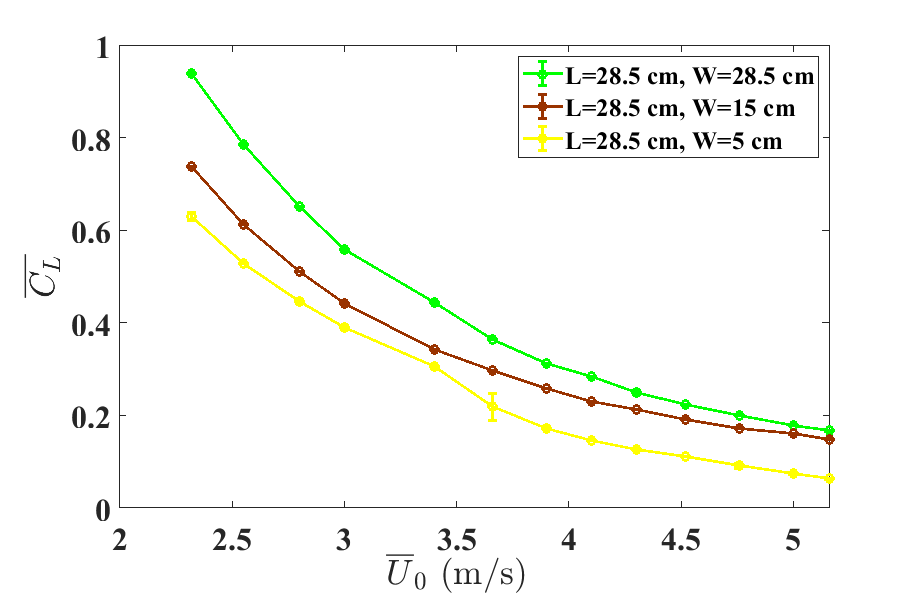}
\caption{ }
\label{fig:Lift coeff_summary for all widths}
\end{subfigure}
\begin{subfigure}[b]{0.49\textwidth}
\centering
\includegraphics[width=\textwidth,trim={0.2cm 0cm 1cm 0},clip]{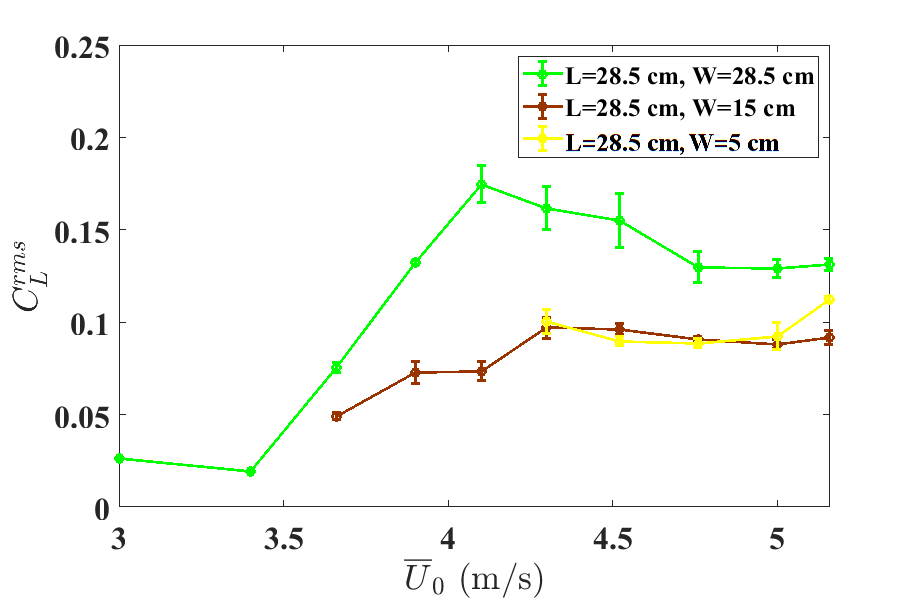}
\caption{ }
\label{fig:Lift coeff_summary for all widths}
\end{subfigure}
\caption{(a) Time history of $C_L$ at $\overline{U}_0 = 4.52$ m/s (b) Mean Lift variation ($\overline{C}_L$) and (c) Rms Lift variation (${C_Lrms}$) with $\overline{U_0}$ for $\AR \in \{1, 0.7, 0.17\}$.}\label{forceAspectRation}
\end{figure}


\begin{figure}
\begin{subfigure}[b]{0.49\textwidth}
\centering
\includegraphics[width=\textwidth,trim={0.75cm 0cm 1cm 0},clip]{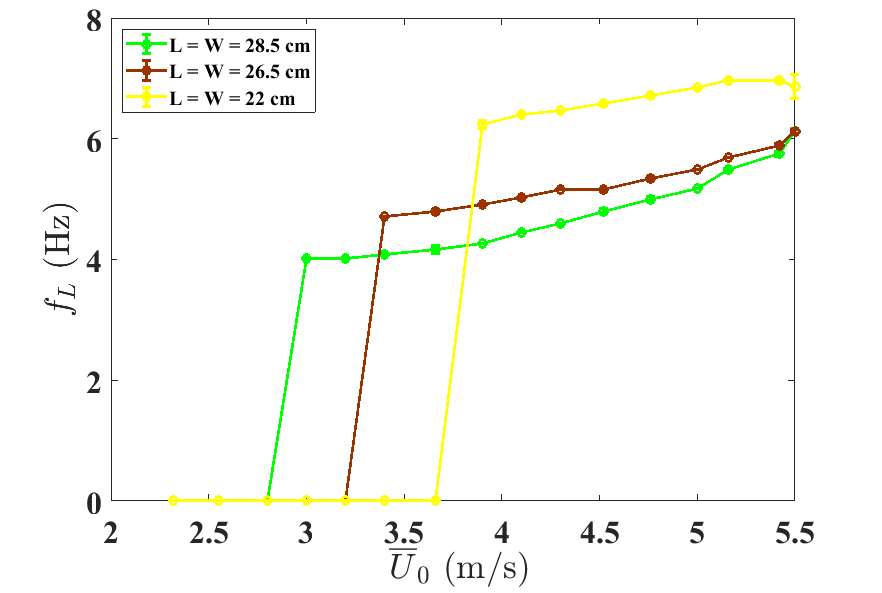}
\caption{ }
\label{fig:forcefreq for all lengths}
\end{subfigure}
\hfill
\begin{subfigure}[b]{0.49\textwidth}
\centering
\includegraphics[width=\textwidth,trim={0.75cm 0cm 1cm 0},clip]{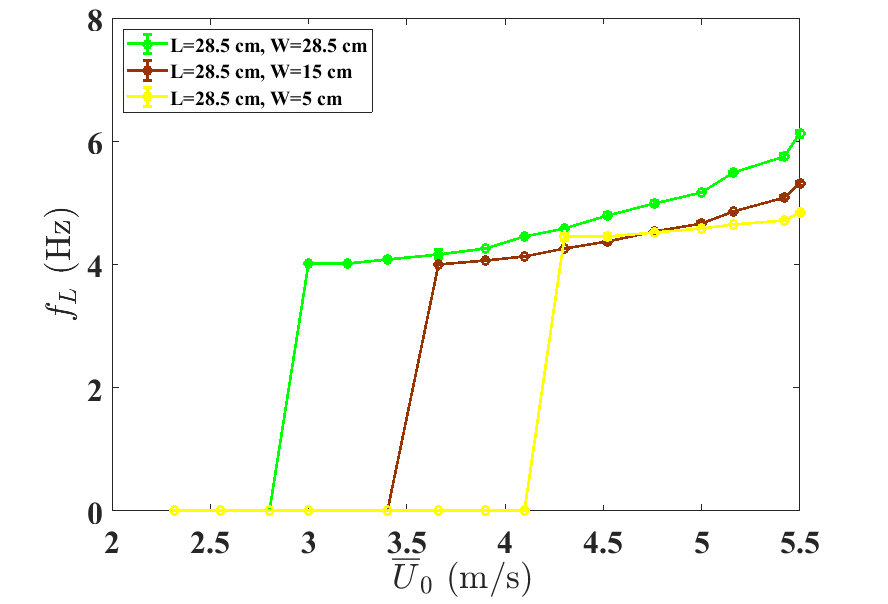}
\caption{ }
\label{fig:forcefreq for all widths}
\end{subfigure}
\caption{ Force Frequency spectrum analysis for (a) $\AR = 1$ (b) $\AR = 0.7$ at $\overline{U_0} = 4.52$ m/s and variation of the force frequencies ($f$) generated for (d) $L\in\{22-28]$ cm for $\AR=1$ (e) $\AR\in\{0.17-1]$ for $L=28.5$~cm. }
\end{figure}
Having discussed the effect of gravity acting along the direction of bending on the kinematics, let us now turn our attention to the lift force dynamics that contributes to the flapping motion of the flexible structure. While the kinematic analysis outlines the motion patterns, examining the corresponding force response helps to complete the overall picture of the flapping behavior. 

Figure~\ref{forceConstLength} presents the behavior of the lift force due to the coupled interactions of a flexible structure over a range of wind velocities and lengths for a constant $\AR=1$. Figure~\ref{forceConstLength}a compares the time history of the lift coefficient, defined as $C_L= F_L/(0.5 \rho U_0^2 L W)$ where $F_L$ is the lift force in Newtons, for three different lengths at $\overline{U}_0=4.52$ m/s. From the figure, it can be observed that for a given $\overline{U}_0$, there exists a distinct dominant frequency that is identical to the periodic flapping frequency. This is further illustrated in figure~\ref{forceConstLength}b, which shows the fast Fourier transform of the lift force for $L=28.5$ cm. In addition to the dominant frequencies, the power spectral distribution plot shows second and third harmonics.
Figures~\ref{forceConstLength}c and~\ref{forceConstLength}d summarize the variation of the mean $C_L$ ($\overline{C_L}$) and root mean squared $C_L$ ($C_L^{rms}$) about the mean over a range of velocities and lengths, respectively. These figures show that $\overline{C_L}$ is independent of of length for a given velocity and it decreases with increasing velocities. The mean lift force corresponds to the weight of the flexible structure and is a constant for a given structure. Hence, $\overline{C_L}$
increases quadratically as the velocity decreases. $C_L^{rms}$
values remain negligibly small during the pre-onset region. Once the flexible structure loses its stability,
$C_L^{rms}$ curves exhibit a sharp increase, can be observed in the figure~\ref{forceConstLength}d. Similar trend in $C_L^{rms}$ has been observed in the numerical investigations reported in \cite{gurugubelli2017thesis} even when gravitational effects are neglected. Unlike the observations reported in \cite{gurugubelli2017thesis}, the drop in $C_L^{rms}$ after the sharp peak is not characterized by a mode change. Instead, the maximum flapping position of TE surpasses the LE mounting level and the drag force component begins to act in a direction opposite to the lift, thus stabilizing the flapping amplitude reported in Section~\ref{kinematics} and leading to a drop in the values of $C_L^{rms}$. 
  


Figure~\ref{forceAspectRation} summarizes the effect of $\AR$ on $C_L$ as a function of wind velocities. Figure~\ref{forceAspectRation}a compares the $C_L$ time histories for three different $\AR$ corresponding to $U_0=4.52$ m/s. It can be clearly observed from the figure that $C_L$ amplitude decreases and $\overline{C_L}$ increases with decreasing $\AR$. This observation is consistent with the ones observed in kinematics, Figures~\ref{forceAspectRation}b and \ref{forceAspectRation}c present the variation of $\overline{C_L}$ and ${C_L^{rms}}$ about the mean for 3 $\AR$ and over a range of velocities. 
For a given $\overline{U}_0$,  $\overline{C}_L$ reduces as $\AR$ reduces due to the reduction in the pressure difference between the top and bottom surfaces because the effect of side vortices become predominant for lower $\AR$. Unlike the variation of $C_L^{rms}$ with velocity for $\AR=1$ in figure~\ref{forceConstLength}d and figure~\ref{forceAspectRation}d, for lower $\AR$ the $C_L^{rms}$ does not exhibit a sudden drop in values after the initial sharp jump indicating the onset of flapping. This observation arises from the flapping motion of the flexible structure largely staying below the mean position of the LE’s mounting point due to its own weight.  

The frequency of the lift forces ($f_L$) obtained is also analyzed for the various $L$ and $\AR$ as shown in figures~\ref{fig:forcefreq for all lengths} and \ref{fig:forcefreq for all widths}. The $f_L$ values follow a similar trend as that of the flapping frequency (f) of the flexible structure, as shown in figure \ref{fig:freq w.r.t length} and \ref{fig:freq w.r.t width}. The similarity in the frequency values shows that the aerodynamic lift for the corresponding flapping kinematics has been adequately captured, as it is consistent with the kinematics of the flexible structure. 

\subsection{Comparative analysis between horizontal and vertical configurations}


\begin{figure}
\centering
\begin{subfigure}[b]{0.49\textwidth}
\centering
\includegraphics[width=\textwidth,trim={0.5cm 0cm 2cm 0},clip]{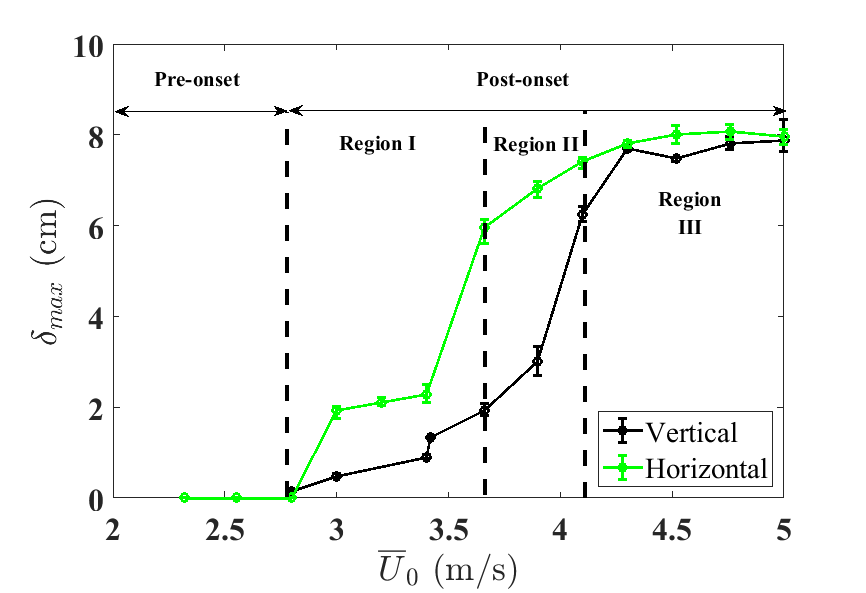}
\caption{ }
\label{fig:Vertical_maxamp}
\end{subfigure}
\hfill
\begin{subfigure}[b]{0.49\textwidth}
\centering
\includegraphics[width=\textwidth,trim={0.5cm 0cm 2cm 0},clip]{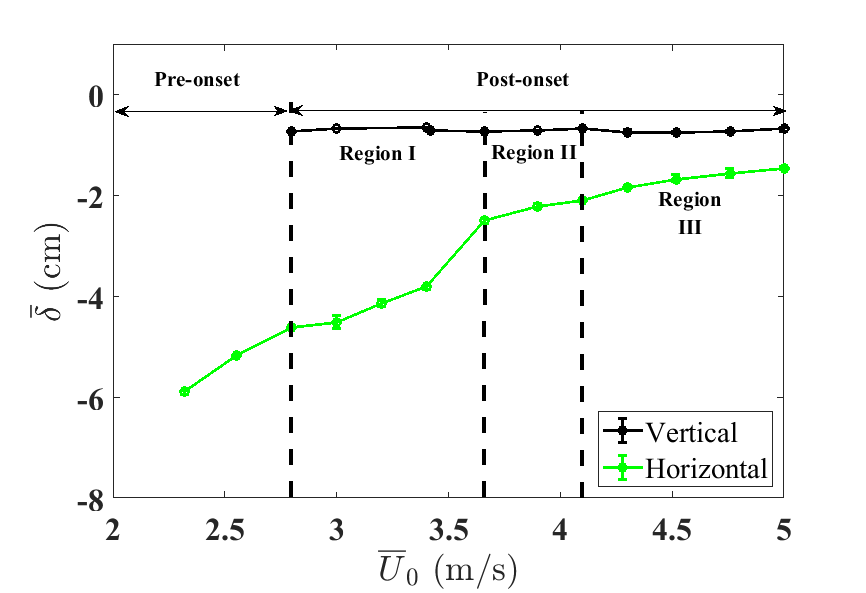}
\caption{ }
\label{fig:Vertical_meanamp}
\end{subfigure}
\begin{subfigure}[b]{0.49\textwidth}
\centering
\includegraphics[width=\textwidth,trim={1cm 1mm 2cm 0},clip]{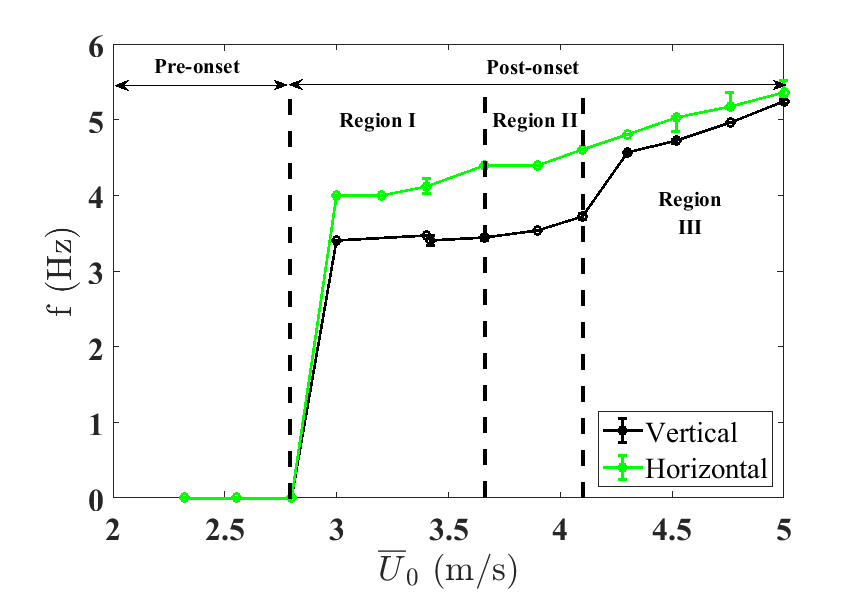}
\caption{ }\label{fig:Vertical_freq}
\end{subfigure}
\caption{ Comparison of the (a) maximum tip displacement ($\delta_{max}$), (b) mean tip displacement ($\bar{\delta}$), and (c) dominant flapping frequency (f) for the vertical and horizontal configurations for $\AR = 1$ and $L=28.5$ cm.}
\end{figure}

\begin{figure}
\centering

\end{figure}

\begin{figure}
\centering
\begin{subfigure}[b]{0.49\textwidth}
\centering
\includegraphics[width=\textwidth,trim={0.02cm 0cm 2cm 0},clip]{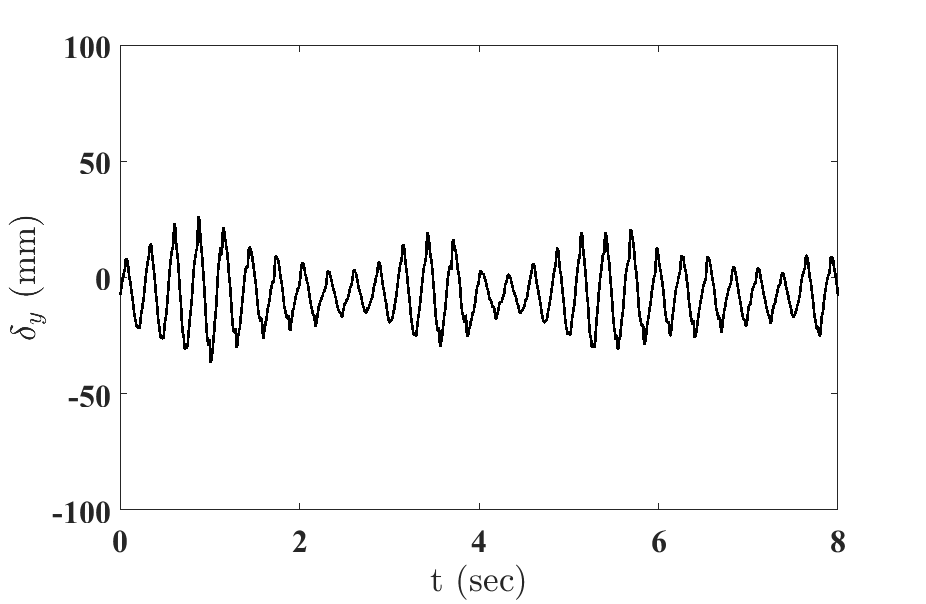}
\caption{ }
\label{fig:Vertical_V3.9_tipdisp}
\end{subfigure}
\hfill
\begin{subfigure}[b]{0.49\textwidth}
\centering
\includegraphics[width=\textwidth,trim={0.02cm 0cm 1cm 0},clip]{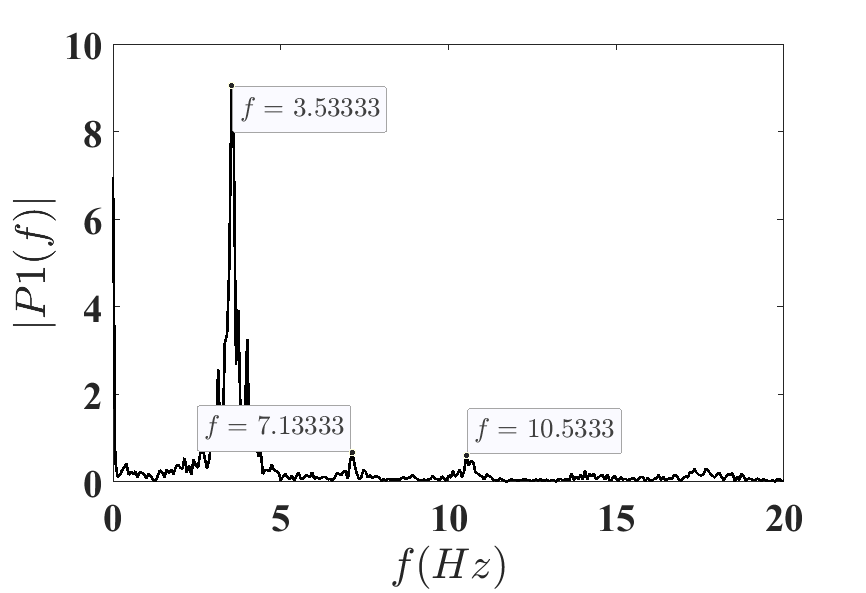}
\caption{ }
\label{fig:Vertical_V3.9_fft}
\end{subfigure}
\caption{ (a) Time history of the transverse tip displacements ($\delta_{y}$) and (b) frequency power spectrum for a vertical sheet of $\AR = 1$ and $L = 28.5$ cm corresponding to Region-II at $\overline{U}_0 = 3.9$ m/s. }
\end{figure}

To further understand the role of gravity along the bending on the flapping instability and response kinematics, wind tunnel experiments have been performed on a vertically mounted flexible sheet. Figures \ref{fig:Vertical_maxamp} to \ref{fig:Vertical_freq} compares the maximum $(\delta_{max})$, mean ($\bar{\delta}$), and dominant frequency (f) of TE flapping response corresponding to both the configurations for a flexible sheet of $\AR=1$ and $L=28.5$ cm. For the pre-onset region, i.e. for velocities $\overline{U}_0\lessapprox3$ m/s, the sheet remains in a stable state for both the configurations, and the flapping instability largely remains unaffected by the gravity. 
The post-critical flapping region can be broadly classified into three regions based on the differences in the flapping kinematics observed between the horizontal and vertical configurations. In the Region-I for $\overline{U}_0 \in [3-3.66]$ m/s, even though both the vertical and horizontal sheets exhibit periodic flapping response, the vertical sheet exhibits significantly lower flapping amplitude in comparison to its horizontal counterpart. Unlike the horizontally mounted flexible structure that undergoes flapping about the deformed state due to its own weight, the vertically mounted structure undergoes flapping about a very small constant mean position or near-zero mean position. Additionally, the flapping frequency for the vertical configuration is relatively lower compared to its horizontal counterpart for Region I. The maximum flapping amplitudes exhibit a sharp jump for the vertical configuration in Region-II, such that the maximum flapping amplitudes and flapping frequencies for both configurations become very close in Region-III. Region-II represents a transition region for the vertical configuration, wherein the TE exhibits an aperiodic flapping response with multiple frequencies corresponding to the higher harmonics. Figures~\ref{fig:Vertical_V3.9_tipdisp} and \ref{fig:Vertical_V3.9_fft} present the time history of transverse tip displacements ($\delta_{y}$) and the frequency power-spectrum for the vertical sheet at $\overline{U}_0 = 3.9$ m/s corresponding to the Region-II that exhibits an aperiodic flapping. The figures clearly show the presence of the second and third harmonics. In this subsection, the investigation is predominantly limited to $\AR=1$ for $L=28.5$ cm sheet. 

Wind tunnel experiments have also been performed for the vertical configuration for $\AR=0.7$ and $0.17$ with $L=28.5$ cm sheet and experiments have shown that the onset remained unaffected by the configurations. However, with reducing $\AR$ the flexible structure in vertical configuration exhibits three-dimensional longitudinal twisting and this twist becomes stronger with reducing $\AR$. The bending and twisting experienced by the vertical sheets with lower $\AR$ contributes to the complex three-dimensional interactions. However, a detailed investigation of these three-dimensional effects falls outside the scope of the current study, which focuses primarily on two-dimensional flapping behavior. 

\section{Conclusion}

In this work, wind tunnel experiments have been performed on a thin flexible structure for two different configurations: (i) horizontally mounted with gravity acting along the direction of bending and (ii) vertically mounted with gravity along the span of the structure, to understand the impact of gravity on the onset of flapping instability and the associated post-critical flapping kinematics and force-dynamic. In this study, a thin mylar sheet is investigated for a range of geometric parameters, i.e. $L\in\{28.5,26.6,22\}$ for $\AR=1$ and $L=28.5$ cm for $\AR\in[0.17-1.07]$, for wind speed up to 5.5 m/s for the horizontal configuration so that the flapping kinematics remains largely two-dimensional. The veritical configuration is investigated for three different $\AR=\{0.17, 0.7,1\}$ at $L=28.5$ cm. The experiments reveal that the gravity does not influence the onset of the flapping instability. The onset velocities remained the same for all three $\AR$. However, it has been observed that a horizontally mounted flexible structure exhibits relatively higher post-critical flapping amplitudes and frequency when compared to their equivalent vertical counterparts. Additionally, because of the weight acting along the bending direction in the case of horizontally mounted configuration, a flexible structure performs a periodic flapping motion about the deformed state with a non-zero mean position. The mean-position of the deformed structure moves closer to zero position, i.e. the LE's mounting level, as the wind velocity increases. Similar to earlier studies neglecting gravitational effects, reduction in $L$ or $\AR$ delay's the onset of instability. The flapping amplitudes and frequency for a shorter structure is lower for a given velocity due to the increase bending stiffness. Similarly, the flapping amplitudes and frequency decrease with $\AR$ due to the effect of side vortices.  
The lift force dynamics exhibits multi-frequency response consisting of higher harmonics even when the flexible structure's TE exhibited a perfectly sinusoidal response. The dominant lift force frequency synchronized with the flapping frequency for all the geometric parameters. The observations presented above are of importance because a flexible structure mounted horizontally exhibits relatively higher flapping amplitudes and frequencies compared to the vertical counterpart for applications in energy harvesting. Additionally, the horizontal configuration offers a better insights in to the development of flexible flapping-wing based propulsive devices due to closer resemblance to bird wing configuration.

\bibliography{cas-refs}

\end{document}